\documentclass[
 reprint,
 amsmath,amssymb,
 aps,
]{revtex4-2}

\usepackage{dcolumn}
\usepackage{bm}
\usepackage[pdftex]{graphicx,color}
\usepackage{comment}
\usepackage{breakcites}
\usepackage{here}
\usepackage[normalem]{ulem}
\usepackage[breaklinks=true]{hyperref}
\usepackage{braket}
\usepackage{url}

\definecolor{myblue}{RGB}{50,50,200}

\begin{document}

\preprint{APS/123-QED}

\title{
Determinant-free fermionic wave function using feed-forward neural networks
}

\author{Koji Inui, Yasuyuki Kato, and Yukitoshi Motome}
\affiliation{
 Department of Applied Physics, The University of Tokyo, Hongo, Tokyo 113-8656, Japan
}

\begin{abstract}
We propose a general framework for finding the ground state of many-body fermionic systems by using feed-forward neural networks. 
The anticommutation relation for fermions is usually implemented to a variational wave function by the Slater determinant (or Pfaffian), which is a computational bottleneck because of the numerical cost of $O(N^3)$ for $N$ particles. 
We bypass this bottleneck by explicitly calculating the sign changes 
associated with particle exchanges in real space and using fully connected neural networks for optimizing the rest parts of the wave function. 
This reduces the computational cost to $O(N^2)$ or less. 
We show that the accuracy of the approximation can be improved by  optimizing the ``variance'' of the energy simultaneously with the energy itself. 
We also find that a reweighting method in Monte Carlo sampling can stabilize the calculation. 
These improvements can be applied to other approaches based on variational Monte Carlo methods. 
Moreover, we show that the accuracy can be further improved by using the symmetry of the system, the representative states, 
and an additional neural network implementing a generalized Gutzwiller-Jastrow factor. 
We demonstrate the efficiency of the method by applying it to a two-dimensional Hubbard model.
\end{abstract}

\pacs{Valid PACS appear here}

\maketitle

\section{Introduction}
\label{introduction}
Recent developments in machine learning have a profound impact on the field of physics~\cite{RevModPhys.91.045002}.
In particular, neural networks (NNs), which in principle can approximate arbitrary functions~\cite{Hornik_Stinchcombe_White_1989,Cybenko_1989}, 
have been intensively studied as a new tool to represent quantum many-body wave functions. 
So far, various methods have been proposed in solid state physics, 
such as the restricted Boltzmann machine~\cite{10.1126/science.aag2302,10.1103/physrevb.96.205152,Choo_Carleo_Regnault_Neupert_2018,10.1038/s42005-021-00609-0}, 
the convolutional neural network (CNN)~\cite{10.1103/physrevb.100.125124,Choo_Carleo_Regnault_Neupert_2018,10.1103/physrevb.103.035138,10.1103/physrevresearch.2.012039,10.1103/physrevresearch.2.033075,10.1103/physrevb.102.205122}, 
the graph convolutional network (GCN)~\cite{10.1103/physrevb.100.125124}, 
the Gaussian process state~\cite{10.1103/physrevx.10.041026}, 
and the fully connected neural network (FCNN)~\cite{10.1103/physrevlett.122.226401}. 
Similar methods have also been applied to molecules in quantum chemistry calculations~\cite{10.1103/physrevresearch.2.033429,10.1038/s41467-020-15724-9,10.1038/s41557-020-0544-y}.

While such studies using NNs are successful for bosonic systems~\cite{10.7566/jpsj.86.093001,z78}, they have encountered some difficulties for fermionic systems and frustrated spin systems due to the complicated sign structures 
in the wave functions~\cite{10.1038/s41467-020-15402-w}. 
For instance, it was reported that the many-variable variational Monte Carlo (mVMC) calculation can achieve higher accuracy compared to NN-based ones for highly-frustrated spin systems~\cite{astrakhantsev2021broken}.
In addition, in most cases the sign structure of the wave function is implemented by the Slater determinant (or Pfaffian) and the rest part of the wave function is approximated by NNs~\cite{10.1103/physrevb.96.205152,10.1103/physrevlett.122.226401,10.1103/physrevb.102.205122}, 
but it is computationally expensive to extend such approaches to large system sizes since the numerical cost for the calculation of the Slater determinant is $O(N^3)$ for $N$-particle systems.
Most recently, methods which do not rely on the Slater determinant have been developed~\cite{10.1038/s41467-020-15724-9,10.1038/s42005-021-00609-0}.

In this paper, we propose a general framework to approximate fermionic many-body wave functions by NNs without using the Slater determinant. 
In our method, the sign changes of the wave function are explicitly calculated for each particle exchange in real space, while the rest part is approximated by the NNs.
This reduces the numerical cost from $O(N^3)$ to $O(N^2)$ or less. 
In addition, we successfully improve and stabilize the calculations by introducing the following developments.
(i) We optimize not only the energy but also the ``variance'' of the energy simultaneously.
(ii) We use a reweighting technique with a modified Monte Carlo (MC) weight to stabilize the calculations of the local energy. 
(iii) We take into account only the ``representative'' states among the symmetry-related ones, which allows us to use flexible NN architectures; we employ the FCNN in the present study.
(iv) We incorporate not only the number of the particles on each degrees of freedom  but also their products as the inputs to the NN.
(v) We prepare an additional NN that is a generalization of the Gutzwiller-Jastrow (GJ) factor. 
We apply the above framework to the Hubbard model on a two-dimensional square lattice, 
and demonstrate that the numerical accuracy can be better than the mVMC result for the modest number of electrons.

This paper is organized as follows. 
In Sec.~\ref{sec:modelandmethod}, we introduce the method. 
After describing the overall framework in Sec.~\ref{subsec:overallframework}, we introduce each component in the framework one by one from Sec.~\ref{subsec:mcmc} to Sec.~\ref{subsec:update}. 
We also discuss the numerical cost in Sec.~\ref{subsec:computationalcost}.
In Sec.~\ref{sec:results}, we present the benchmark results for the Hubbard model. 
After describing the set up of the calculations in Sec.~\ref{subsec:setup}, 
we demonstrate the efficiency of our technique by changing 
the parameters in the model and method in Secs.~\ref{subsec:res_mcmc}-\ref{subsec:6x6}.
Finally, we give the concluding remarks in Sec.~\ref{sec:concludingremarks}.

\begin{figure*}[htbp]
        \centering
        \includegraphics[width=1.9\columnwidth,pagebox=cropbox,clip]{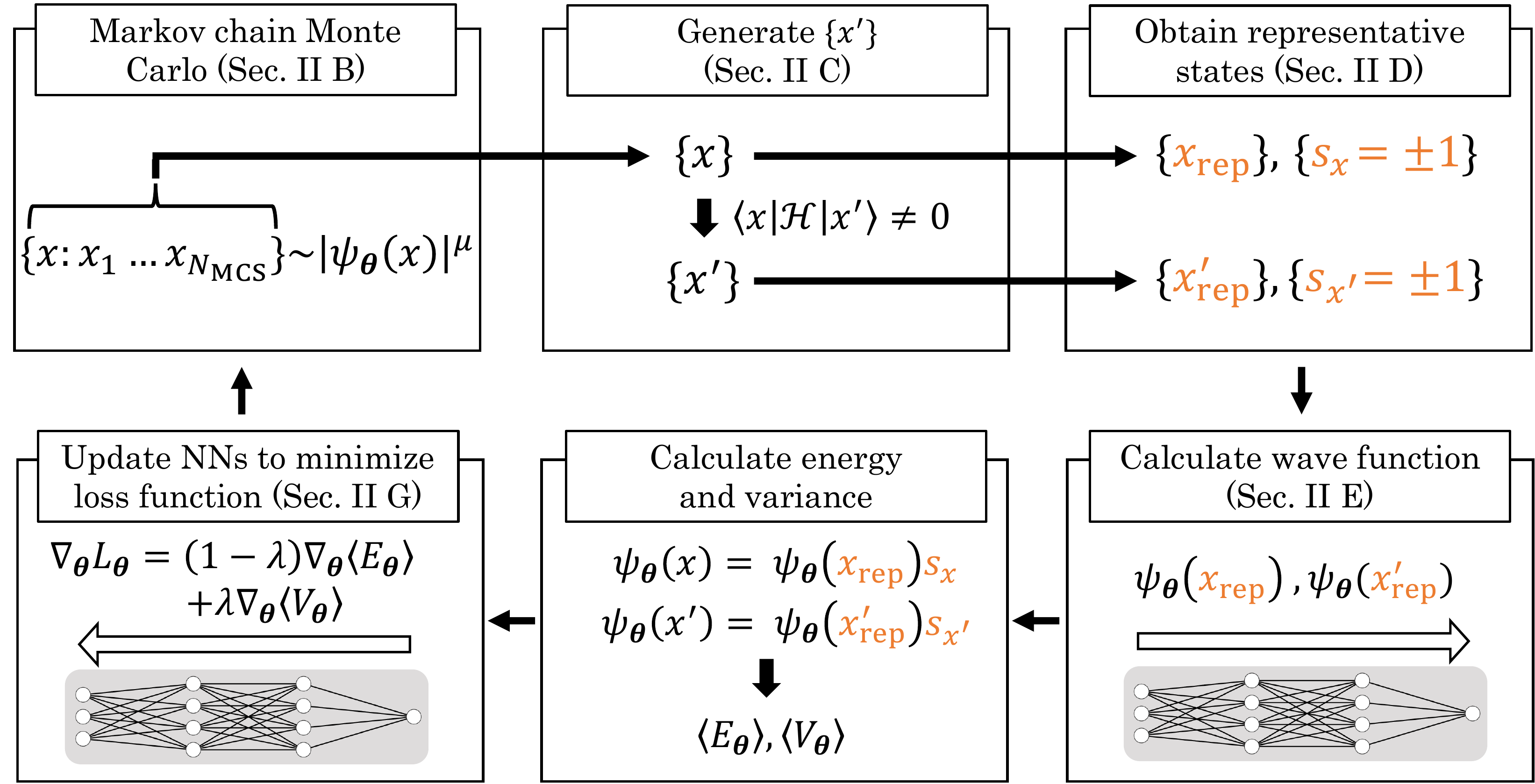}
        \caption{
        Flowchart of the present calculation. 
        For details of each part, refer to the sections indicated in the figure.
}
        \label{Fig:flow}
\end{figure*}

\section{Method}
\label{sec:modelandmethod}

\subsection{Overall framework}
\label{subsec:overallframework}

In this paper, we present our framework for discrete lattice systems, 
while the method is generic and can be straightforwardly extended to the systems in continuous space like an electron gas and molecules.
First, we denote the state of $N$ fermionic particles on a lattice as
\begin{equation}
| x \rangle = |(\bm{r}_{1}, \sigma_1), (\bm{r}_{2}, \sigma_2), (\bm{r}_{3}, \sigma_3), \ldots, (\bm{r}_N, \sigma_N) \rangle,
\label{eq:x}
\end{equation}
where $\bm{r}_{i}$ and $\sigma_i$ denote the spatial coordinates and the internal degrees of freedom such as spin of $i$th particle, respectively.
In Eq.~\eqref{eq:x}, $\{(\bm{r}_i, \sigma_i)\}$ are arranged in ascending order of an integer $I(\bm{r}, \sigma) \in [0,N_{\rm tot})$ which labels the one-particle states $(\bm{r},\sigma)$; 
$N_{\rm tot}=N_{\rm site} N_{\rm int}$, where $N_{\rm site}$ and $N_{\rm int}$ are the numbers of the lattice sites and the internal degrees of freedom, respectively.

For a given Hamiltonian $\mathcal{H}$, 
the local energy for each state $|x\rangle$ is defined as
\begin{align}
E^{\rm{loc}}_{\bm{\theta}}(x)  = \frac{\langle x |\mathcal{H} |\psi_{\bm{\theta}} \rangle}{\langle x |\psi_{\bm{\theta}} \rangle}
\label{eq:e_loc}
\end{align}
where $|\psi_{\bm \theta}\rangle$ is an arbitrary wave function.
Then, the expectation value of the total energy can be written as 
\begin{align}
\langle E_{\bm{\theta}} \rangle 
&=\frac{ \langle \psi_{\bm{\theta}} | \mathcal{H} | \psi_{\bm{\theta}} \rangle }{
 \langle \psi_{\bm{\theta}}  | \psi_{\bm{\theta}} \rangle } \nonumber \\
&= \sum_x \frac{|\langle x|\psi_{\bm{\theta}}\rangle|^2}{\langle\psi_{\bm{\theta}}|\psi_{\bm{\theta}}\rangle}\frac{\langle x |\mathcal{H} |\psi_{\bm{\theta}} \rangle}{\langle x |\psi_{\bm{\theta}} \rangle} \nonumber \\
&= \sum_x p_{\bm{\theta}}(x) E^{\rm{loc}}_{\bm{\theta}}(x),
\label{eq:E}
\end{align}
where $p_{\bm{\theta}}(x)$ is the probability distribution of the state $|x\rangle$ given by 
\begin{align}
&p_{\bm{\theta}}(x) = \frac{|\psi_{\bm{\theta}}(x)|^2}{\sum_{x'}|\psi_{\bm{\theta}}(x')|^2},\\
&\psi_{\bm{\theta}}(x) = \langle x | \psi_{\bm{\theta}} \rangle.
\label{fig:psi}
\end{align}
From the variational principle, the relation 
$\langle E_{\bm{\theta}} \rangle \geq E_{\rm{GS}}$
always holds, where $E_{\rm GS}$ is the true ground state energy. 
In addition, we define the ``variance" of the energy by 
\begin{align}
\langle V_{\bm{\theta}} \rangle= \sum_x p_{\bm{\theta}}(x) V_{\bm{\theta}}^{\rm{loc}},
\label{eq:V}
\end{align}
with
\begin{align}
V_{\bm{\theta}}^{\rm{loc}} =  
|E^{\rm{loc}}_{\bm{\theta}}(x) - \langle E_{\bm{\theta}} \rangle |, 
\label{eq:V_loc}
\end{align}
which satisfies the relation 
$\langle V_{\bm{\theta}} \rangle\geq 0$
since all the local energies in the true ground state are equal to $E_{\rm GS}$.

In this framework, we approximate the wave function $\psi_{\bm{\theta}}(x)$ in Eq.~\eqref{fig:psi} by
using NNs with the parameters $\bm{\theta}$, and optimize it so as to minimize not only the energy $\langle E_{\bm{\theta}} \rangle$ but also $\langle V_{\bm{\theta}} \rangle$ simultaneously; namely, we minimize the loss function defined as 
\begin{align}
L_{\bm \theta} = (1-\lambda) \langle E_{\bm{\theta}} \rangle +  \lambda \langle V_{\bm{\theta}} \rangle,
\label{eq:L}
\end{align}
where $\lambda$ is a hyperparameter that we tune during the optimization process ($0 \leq\lambda <1$).

The overall framework of our method is shown in Fig.~\ref{Fig:flow}.
First, we generate many states $\{|x\rangle\}$ by using the Markov chain MC (MCMC) sampling (Sec.~\ref{subsec:mcmc}).
Next, we generate the states $\{|x^\prime \rangle\}$ for $\{|x\rangle\}$, satisfying 
$\langle x | \mathcal{H} | x^{\prime} \rangle \neq 0$, which are necessary to calculate the local energy $E^{\rm{loc}}_{\bm{\theta}}(x)$ in Eq.~\eqref{eq:e_loc} (Sec.~\ref{subsec:genx}).
Then, we reduce them to the representative states $\{|x_{\rm rep}\rangle\}$ and $\{|x^{\prime}_{\rm rep}\rangle\}$ 
through the symmetry operations allowed in the given system, and at the same time,  
we calculate the sign changes $\{s_x\}$ and $\{s_{x^\prime}\}$ caused by particle exchanges under the symmetry operations (Sec.~\ref{subsec:symm}).
In the next step, we calculate $\{\psi_{\bm{\theta}}(x_{\rm rep})\}$ and $\{\psi_{\bm{\theta}}(x_{\rm rep}^{\prime})\}$ 
by using the NNs composed of two FCNNs with different roles (Sec.~\ref{subsec:nn}). 
Then, we calculate $\langle E_{\bm \theta} \rangle$ and $\langle V_{\bm \theta} \rangle$ 
by using $\psi_{\bm{\theta}}(x) = \psi_{\bm{\theta}}(x_{\rm rep}) s_{x}$ and 
$\psi_{\bm{\theta}}(x^{
\prime}) = \psi_{\bm{\theta}}(x^{
\prime}_{\rm{repr}}) s_{x^{
\prime}}$. 
Finally, we compute the derivatives of the loss function in terms of $\bm{\theta}$, $\nabla_{\bm{\theta}}L_{\bm \theta}$, through the backpropagation, and update the parameters $\bm \theta$ (Sec.~\ref{subsec:update}). 
The updated NNs are used to generate new states $\{|x\rangle\}$ in the next loop.
This sequence is repeated until the energy converges.

\subsection{Markov chain Monte Carlo sampling and reweighting}
\label{subsec:mcmc}

In the conventional variational MC method, 
the states $\{|x\rangle\}$ are generated by the MCMC sampling by using $|\psi_{\bm{\theta}} (x)|^2$ as the weight, 
and the energy is calculated by
\begin{align}
\langle E_{\bm{\theta}} \rangle \approx \frac{1}{N_{\rm{MCS}}} \sum_{\{|x\rangle\}\sim |\psi_{\bm{\theta}}(x)|^2} E^{\rm{loc}}_{\bm{\theta}}(x), 
\label{eq:conventionalMCMC}
\end{align}
where $N_{\rm{MCS}}$ is the number of the MCMC samples.
In this case, however, when a sample has a small value of $|\psi_{\bm{\theta}}(x)|$, 
$E_{\bm{\theta}}^{\rm{loc}}(x)$ in Eq.~\eqref{eq:e_loc} becomes large, which can make the calculation unstable.
This indeed occurs especially in fermionic systems, where $\psi_{\bm{\theta}}(x)$ changes its sign across zero 
during the optimization process.
To circumvent this instability, we perform the MCMC sampling by using $|\psi_{\bm{\theta}}(x)|^\mu$ ($0<\mu\leq1$) as the weight and adopt the reweighting method; namely, we calculate the energy as
\begin{align}
\langle E_{\bm{\theta}} \rangle  \approx
\frac{  \sum_{\{|x\rangle\} \sim |\psi_{\bm{\theta}}(x)|^\mu} |\psi_{\bm{\theta}}(x)|^{(2-\mu)}E^{\rm{loc}}_{\bm{\theta}} (x)}{\sum_{\{|x\rangle\} \sim|\psi_{\bm{\theta}}(x)|^\mu} |\psi_{\bm{\theta}}(x)|^{(2-\mu)}}. 
\label{eq:e_mu}
\end{align}
This trick stabilizes the calculation as 
$|\psi_{\bm{\theta}}(x)|^{(2-\mu)}E^{\rm{loc}}_{\bm{\theta}} (x) = |\psi_{\bm{\theta}}(x)|^{(1-\mu)} \langle x |\mathcal{H} |\psi_{\bm{\theta}} \rangle$ 
with $1-\mu\geq 0$.
We employ the Metropolis algotirhm for the MCMC sampling.

\subsection{Generation of $|x^{\prime}\rangle$}
\label{subsec:genx}

The calculation of $E_{\bm \theta}^{\rm loc}(x)$ 
requires the states $\{|x^\prime\rangle\}$ 
for which the matrix elements $\langle x|\mathcal{H}|x^\prime \rangle$ are nonzero as 
\begin{align}
E_{\bm \theta}^{\rm loc}(x) = \sum_{x'} \frac{\langle x | \mathcal{H} | x' \rangle \langle x' | \psi_{\bm \theta} \rangle}{\langle x | \psi_{\bm \theta} \rangle}. 
\label{eq:E_loc2}
\end{align}
We can generate $\{|x^\prime\rangle\}$ by simply operating $\mathcal{H}$ on $|x\rangle$. 
In these operations, however, the state in $|x^\prime\rangle$, in general, violates the order of the particles in Eq.~\eqref{eq:x}. 
In such cases, we need to rearrange the order to follow the representation in Eq.~\eqref{eq:x}. 
The numerical cost is $O(N\log N)$. 
At the same time, the overall sign is changed by $(-1)^P$, where $P$ is the number of times for the particle exchanges in the rearrangement of $|x^\prime\rangle$.
Note that when $|x^\prime\rangle$ is generated by the single-particle hopping, 
the numerical cost is reduced to  $O(\log N)$.
In this case, $P$ is given by the change in the order of the particle associated with the hopping.

\subsection{Symmetry operation and representative state}
\label{subsec:symm}

The wave functions for the states connected by the symmetry of the system, 
such as the point group and translational symmetries, are identical except for the phases.
Therefore, it is sufficient to take into account only one of such equivalent states in the actual calculations. 
In order to distinguish the equivalent states, we introduce 
\begin{align}
    m_x = \sum_{(\bm{r},\sigma) \in x} 2^{N_{\rm tot}-{I}(\bm{r},\sigma)}n_{\bm{r} \sigma},
\label{m_x}
\end{align}
where 
$n_{\bm{r} \sigma} = \langle x | \hat{n}_{\bm{r}\sigma} |x \rangle$ 
with 
$\hat{n}_{\bm{r}\sigma} =  \hat{c}_{{\bm{r}}\sigma}^\dagger \hat{c}_{{\bm{r}}\sigma}$;  
$\hat{c}_{{\bm{r}}\sigma}^\dagger$ ($\hat{c}_{{\bm{r}}\sigma}$)  is the creation (annihilation) operator of the fermion with ${\bm{r}}$ and $\sigma$.
We call the state with the largest value of $m_x$ the ``representative state" 
among the symmetry-related ones, and denote it as $| x_{\rm rep} \rangle$. 
For each $|x\rangle$, we retain the sign change $s_x$ from $| x \rangle$ to $| x_{\rm rep} \rangle$, which consists of 
the product of the sign changes caused by the particle exchanges and the character of the symmetry operation.

In the above calculations for the representative states, the symmetry operations in each point group are implemented in a separate manner. 
For example, in the case of $C_{4v}$ symmetry, all the symmetry operations can be constructed by combinations of the fourfold rotation $C_4$ and the mirror operations $\sigma_v$; hence, we implement the algorithms for these two separately. 
The sign change is calculated by multiplying the characters of the symmetry operations according to the irreducible representation of the point group. 
For example, for the irreducible representation $B_2$ in $C_{4v}$, the sign is obtained by multiplying $-1$ ($+1$) for each operation of $C_4$ ($\sigma_v$). 
On the other hand, for the translational symmetry operations, it is sufficient to take into account the state in which 
one of the particles occupies the state with ${I}(\bm{r},\sigma)=0$, since we take into account the representative state with the largest $m_x$ in the calculations.

The bottleneck of the present computation is in the calculations of $m_x$ for all the states generated by the symmetry operations, whose total computational cost is $O(N^2)$. 
In the simulations for finite-size systems, the irreducible representations of the ground state depend on the parameters, such as the system sizes and the number of particles. 
Therefore, we perform the calculations independently for all the irreducible representations 
that do not have degeneracy, 
and select the ground state by comparing the energies among them. 
Note that our framework can also be applicable to the systems without any symmetry, such as low-symmetric molecules and disordered systems.

\subsection{Neural network architecture}
\label{subsec:nn}

\begin{figure}[htbp]
        \centering
        \includegraphics[width=\columnwidth,clip,trim=0 0 0 20]{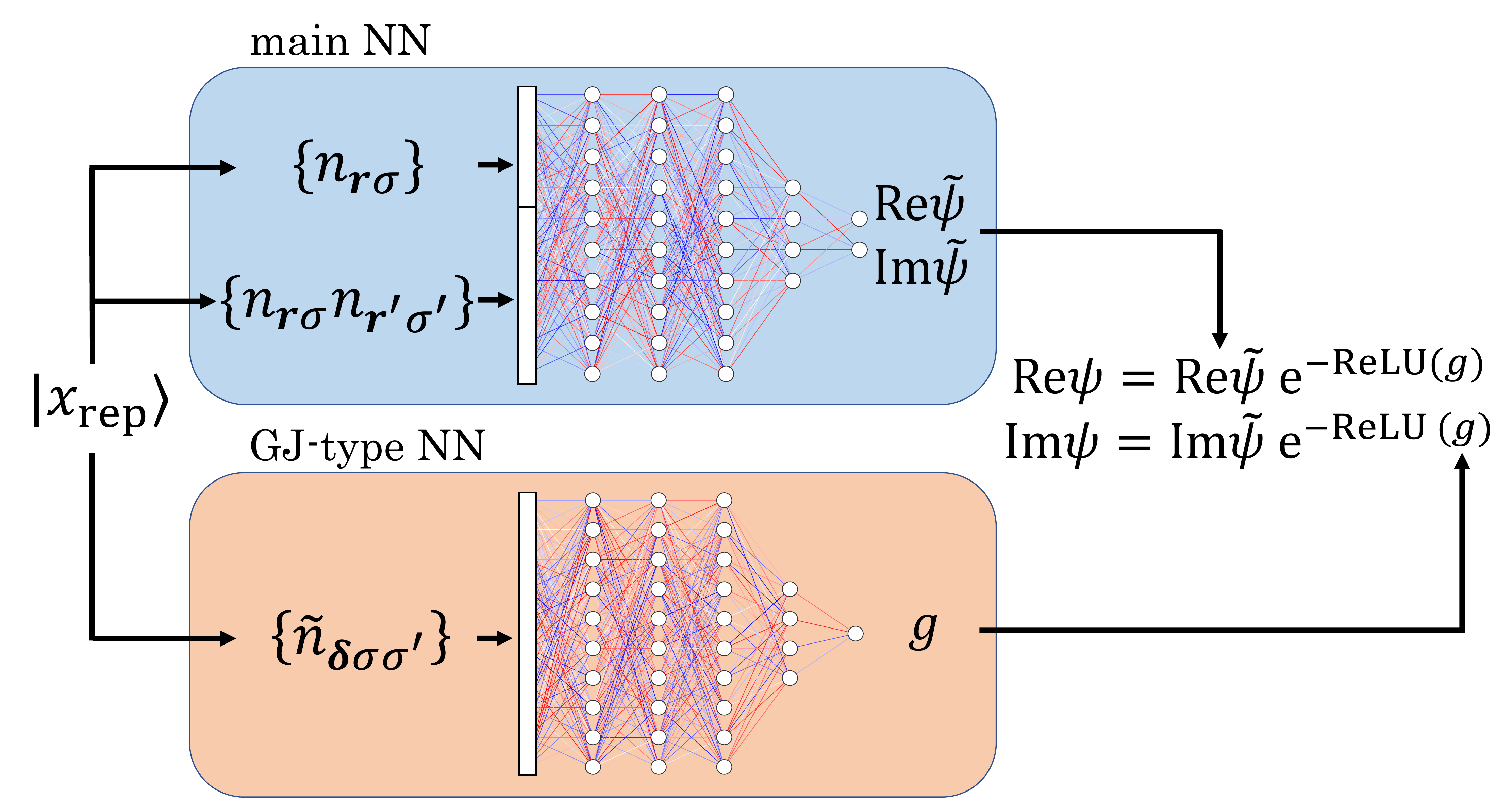}
        \caption{
        NN architecture composed of two types of FCNNs.
        One is called the main NN, which generates the output of two real numbers corresponding to the real and imaginary parts of the wave function 
        from the input of the number of particles at each site, $n_{\bm{r} \sigma}$, and their product $n_{\bm{r} \sigma} n_{\bm{r}^\prime \sigma^\prime}$.
        The other called the GJ-type NN is an extension of the GJ factor, 
        which generates the output of a real number $g$ from the input of the correlation factor for the number of particles in Eq.~\eqref{eq:gj}.
}
        \label{Fig:nn}
\end{figure}

Figure~\ref{Fig:nn} shows the architecture of the NN in this study. 
We use two FCNNs to represent the wave function.
The use of FCNNs brings, at least, two advantages: (i) they can be tuned irrespective of the lattice structure of the system, 
unlike CNNs and GCNs, and 
(ii) they can capture spatially long-range correlations within reasonable computational time and memory consumption.
We tried other architectures, such as Resnet~\cite{7780459} and self-attention network~\cite{vaswani2017attention}, 
but did not find any improvement. 
Below, we describe the details of each FCNN that we used in the present study.

One of the FCNNs, which we call the main NN, is used to represent the real and imaginary parts of the wave function. 
As the input, we use a one-dimensional vector [\{$n_{\bm{r} \sigma}$\},\{$n_{\bm{r} \sigma} n_{\bm{r}^\prime \sigma^\prime}$\}], 
which contains the particle number at each site, $n_{\bm{r} \sigma}$, 
and the products of the particle numbers, $n_{\bm{r} \sigma} n_{\bm{r}^\prime \sigma^\prime}$. 
Since the dimensions of $n_{\bm{r} \sigma}$ and $n_{\bm{r} \sigma} n_{\bm{r}^\prime \sigma^\prime}$ are $N_{\rm tot}$ and $N_{\rm{tot}}(N_{\rm{tot}}-1)/2$, respectively, 
the total dimension of the input vector is $N_{\rm{tot}}(N_{\rm{tot}}+1)/2$.

In order to further take into account the correlation effects, 
we prepare the other FCNN which we call the GJ-type NN. 
The input is the correlation factor for the number of particles defined by
\begin{align}
\tilde{n}_{\bm{\delta}\sigma\sigma^\prime} =  \sum_{\bm{r}} n_{\bm{r}\sigma}n_{\bm{r}+\bm{\delta} \sigma^\prime},
\label{eq:gj}
\end{align}
where $0 \leq \tilde{n}_{\bm{\delta}\sigma \sigma^\prime} \leq N$, and  
the output is a real number $g$, which is incorporated into the wave function as the factor of $e^{-{\rm ReLU}(g)}$, 
as shown in Fig.~\ref{Fig:nn}.
This NN can be regarded as an extension of the GJ factor~\cite{PhysRevLett.10.159,PhysRev.98.1479}.

A unique feature of our NN architecture is that not only the correlation effect but also the sign structure of the fermionic wave function 
is represented by the NN. 
In contrast, most previous studies have used NNs to incorporate the correlation effect, 
with the use of the Slater determinant or the Pfaffian to implement the fermionic anticommutation relations.
Thus, our framework would be more flexible to optimize the sign structure of the wave function than those based on the Slater determinant or the Pfaffian.
Another feature is that not only the number of particles but also the products of them, 
such as $n_{\bm{r} \sigma} n_{\bm{r}^\prime \sigma^\prime}$ and $\tilde{n}_{\bm{\delta}\sigma \sigma'}$, 
are incorporated into the input, which improves the optimization.  
One can regard it as a ``pre-processing'' in the context of machine learning. 
It is easily implemented by employing the FCNNs where the dimension of the input can be tuned flexibly, in contrast to CNNs and GCNs.

\subsection{Update \texorpdfstring{$\bm{\theta}$}{TEXT}}
\label{subsec:update}

The parameters $\bm{\theta}$ are updated so as to minimize the loss function $L_{\bm \theta}$ in Eq.~\eqref{eq:L}.
This is done by calculating the $\bm{\theta}$ derivative of $L_{\bm \theta}$ with using the following equation: 
\begin{widetext}
\begin{align}
\nabla_{\bm{\theta}} \langle O_{\bm{\theta}} \rangle  =
\frac{ 
     \sum_{\{x\} \sim |\psi_{\bm{\theta}}(x)|^\mu} \left[
        ( O^{\rm{loc}}_{\bm{\theta}}(x)-\frac{2}{\mu}\langle O_{\bm{\theta}} \rangle) 
        |\psi_{\bm{\theta}}(x)|^{(2-\mu)} \nabla_{\bm{\theta}} \log (|\psi_{\bm{\theta}}(x)|^\mu) 
        + \nabla_{\bm{\theta}} |\psi_{\bm{\theta}}(x)|^{(2-\mu)}O^{\rm{loc}}_{\bm{\theta}} (x)
    \right]
}{
    \sum_{\{x\} \sim|\psi_{\bm{\theta}}(x)|^\mu} |\psi_{\bm{\theta}}(x)|^{(2-\mu)}
},
\label{eq:o}
\end{align}
\end{widetext}
where $O$ denotes $E$ or $V$;
see Appendix~\ref{app:lf} for the derivation. 
The derivatives in each NN layer are computed by backpropagation.
For the update, we choose an appropriate algorithm, such as the stochastic gradient descent and Adam~\cite{kingma2014adam}, 
depending on the system and the layer architecture of the NN.

\subsection{Computational cost}
\label{subsec:computationalcost}

Let us discuss the computational cost for the calculation of $\psi_{\bm \theta}(x^\prime)$ from $|x^\prime \rangle $, which corresponds to the calculation of the Slater determinant (or Pfaffian) in the previous studies. 
As mentioned in Sec.~\ref{subsec:symm}, 
the bottleneck for preparing $|x^\prime \rangle$ is in the transformation from $|x^\prime \rangle$ to $|x_{\rm rep}^\prime \rangle$, which is $O(N^2)$.
Meanwhile, the preparations of $\{n_{\bm{r}\sigma} n_{\bm{r}^\prime \sigma^\prime}\}$ and 
$\tilde{n}_{\bm{\delta}\sigma\sigma^{\prime}}$ in Sec.~\ref{subsec:nn} both require the cost of $O(N_{\rm tot}^2)$.
Thus, the computational cost in our present scheme is $O(N^2)$. 
This is smaller than the that for the Slater determinant, $O(N^3)$. 
Note, however, that the overall cost in the total optimization depends on the numbers of MCMC samples and NN parameters necessary for each system.

It is possible to further reduce the computational cost in the present method. 
As mentioned in Sec.~\ref{subsec:symm}, the symmetry operations are not mandatory.
One may also omit $\{n_{\bm{r}\sigma} n_{\bm{r}^\prime \sigma^\prime}\}$ in the input for the main NN. 
In addition, the GJ-type NN can be omitted. 
With these simplifications, the bottleneck remains in the generation of $|x^\prime \rangle$, which is $O(N \log N)$. 
As mentioned in Sec.~\ref{subsec:genx}, this is further reduced to $O(\log N)$ 
when $|x^\prime \rangle$ is generated only by single-particle hoppings. 
In this case, the bottleneck is in the calculation of $\{n_{\bm{r}\sigma}\}$ in Sec.~\ref{subsec:nn}, which is $O(N)$.

\section{Results}
\label{sec:results}

In this section, we present the benchmark results by applying our method to the Hubbard model on a square lattice. 
The Hamiltonian reads 
\begin{equation}
\mathcal{H}  = -t \sum_{\langle \bm{r},\bm{r}' \rangle, \sigma}(\hat{c}^\dagger_{\bm{r} \sigma} \hat{c}_{\bm{r}' \sigma} + {\rm h.c.})
+ U\sum_{\bm{r}} \hat{n}_{\bm{r}\uparrow} \hat{n}_{\bm{r}\downarrow}, 
\label{eq:ham}
\end{equation}
where $\langle {\bm r},{\bm r}'\rangle $ denotes the nearest-neighbor pairs, and $\sigma = \uparrow$ or $\downarrow$
($N_{\rm int}=2$). 
We set $t=1$ and $U=4$ throughout the following calculations. 
After describing the computational details in Sec.~\ref{subsec:setup}, 
we first present the benchmark for the system with $N_{\rm site} = 4 \times 4$ and $N=10$ in Secs.~\ref{subsec:res_mcmc}--\ref{subsec:numdep},  
in comparison with the results obtained by the exact diagonalization (ED). 
Then, we show the results for $N_{\rm site}=6\times 6$ and $N=10$ in Sec.~\ref{subsec:6x6}, in comparison with the mVMC result.
In both cases, we impose the periodic boundary conditions, where the noninteracting ground states are in the closed-shell configurations. 
We find that the point group symmetry of the lowest-energy state is always $A_1$ as discussed in Sec.~\ref{subsec:irre}, 
and hence, show the results for $A_1$ in the other sections.

\subsection{Computational details}
\label{subsec:setup}

We perform the MCMC updates for $10^7$-$10^9$ times, and take the samples in every $10^3$-$10^4$ steps, namely, $N_{\rm MCS}=10^4$-$10^5$, except for Sec.~\ref{subsec:numdep}.
In the main (GJ-type) NN, we prepare five (four) layers of the FCNN; 
the first four (three) layers have $400$ parameters, while the last one has $40$. 
In both NNs, we employ the parametric ReLU~\cite{7410480}, except for the last two layers (last layer) for the main (GJ-type) NN where we use the hyperdolic tangent.
We use Adam for the optimization~\cite{kingma2014adam}, in which we take the hyperparameters as $\beta_1=0.9$ and $\beta_2=0.99$. 
For the learning rates $\eta_{\rm main}$ and $\eta_{\rm GJ}$ used in the main and GJ-type NNs, respectively, we set the initial values at 
$10^{-3}$-$10^{-4}$ and gradually decrease them to $10^{-4}$-$10^{-5}$ during the optimization processes; 
the actual schedules will be shown for each case in the following sections.
We note that the optimization process sensitively depends on the values of $\beta_1$, $\beta_2$, 
$\eta_{\rm main}$, and $\eta_{\rm GJ}$. 
Specifically, it is crucial to take $\eta_{\rm GJ}$ being much smaller than $\eta_{\rm main}$ 
to avoid a metastable state where only the onsite Coulomb energy tends to be optimized but the kinetic energy does not.
The calculations are performed on GPUs with parallelization up to 32 cores.

\subsection{Efficiency of the reweighting method}
\label{subsec:res_mcmc}
\begin{figure}[htbp]
        \centering
        \includegraphics[width=\columnwidth]{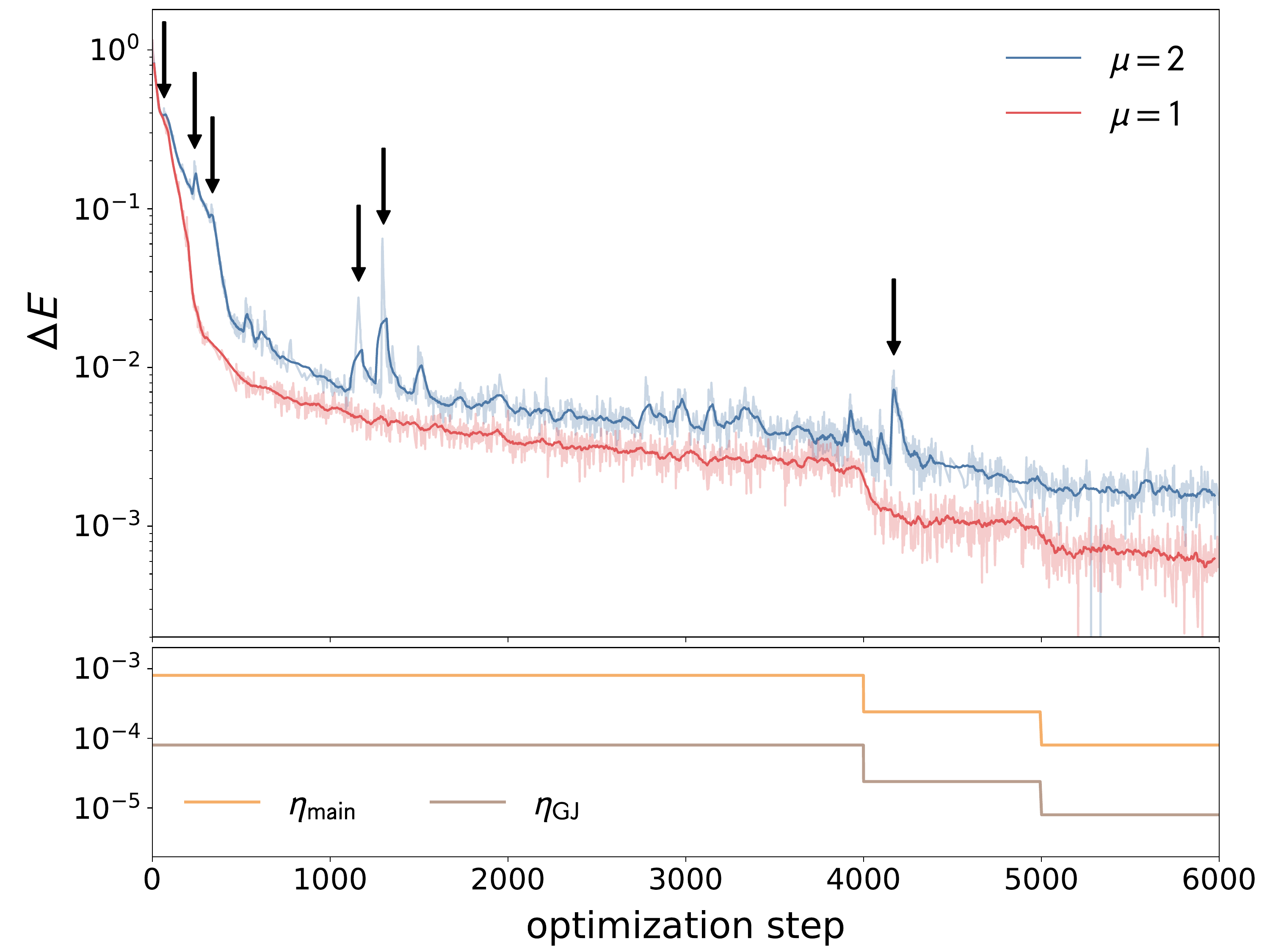}
        \caption{
        Effect of the reweighting method on the optimization process for $N_{\rm site}=4\times 4$ and $N=10$. 
        $\Delta E$ in Eq.~\eqref{eq:DeltaE} is plotted for $\mu=2$ and $\mu=1$ in Eq.~\eqref{eq:e_mu}; 
        the former is equivalent to Eq.~\eqref{eq:conventionalMCMC}. We take $\lambda=0$ in Eq.~\eqref{eq:L}. 
        The dark lines are the moving averages over 20 steps. 
        The arrows for the data for $\mu=2$ indicate the steps where $\Delta E$ suddenly increases because of numerical instabilities. 
        The lower panel shows the schedules of $\eta_{\rm main}$ and $\eta_{\rm GJ}$ commonly used for the calculations with $\mu=2$ and $\mu=1$. 
        }
        \label{Fig:mcmc}
\end{figure}

First, we demonstrate that the reweighting method introduced in Sec.~\ref{subsec:mcmc} can stabilize and accelerate the optimization. 
Figure~\ref{Fig:mcmc} shows the optimization processes of the relative 
error of the energy defined by
\begin{equation}
    \Delta E = \frac{\langle E_{\bm \theta} \rangle - E_{\rm GS}}{|E_{\rm{GS}}|},
    \label{eq:DeltaE}
\end{equation}
when we take $\mu=2$ and $\mu=1$ in Eq.~\eqref{eq:e_mu}. 
Here $E_{\rm GS}$ is the ground state energy obtained by the ED calculation.
We set $\lambda=0$ in Eq.~\eqref{eq:L}. 
In the case of $\mu=2$, which corresponds to the conventional MCMC sampling with the use of Eq.~\eqref{eq:conventionalMCMC}, 
there appear some spikes indicated by the arrows in Fig.~\ref{Fig:mcmc}, 
suggesting numerical instabilities where the local energy becomes large due to small absolute values of the wave function. 
Such behaviors are not observed for the reweighting method with $\mu=1$. 
By this stabilization, the energy for $\mu=1$ is always lower than that for $\mu=2$, 
indicating that the reweighting method accelerates the optimization. 
We also performed the calculations by taking $\mu=0.5$ and $\mu=0.25$, but did not find any further improvement compared to the case of $\mu=1$.

The lower panel of Fig.~\ref{Fig:mcmc} shows the schedules of the learning rates, 
where we reduce $\eta_{\rm main}$ and $\eta_{\rm GJ}$ in a stepwise manner during the optimization. 
We find that this is an efficient way to lower the energy. 
We note that the optimization is no longer accelerated if we take smaller values of 
$\eta_{\rm main}$ and $\eta_{\rm GJ}$ than the smallest ones in Fig.~\ref{Fig:mcmc} 
due to the slowing down of the optimization process. 
Similar behaviors were reported in a previous study~\cite{10.1103/physrevresearch.2.033075}.

\subsection{Two types of the neural networks}
\label{subsec:nrnr}

\begin{figure}[htbp]
        \centering
        \includegraphics[width=\columnwidth]{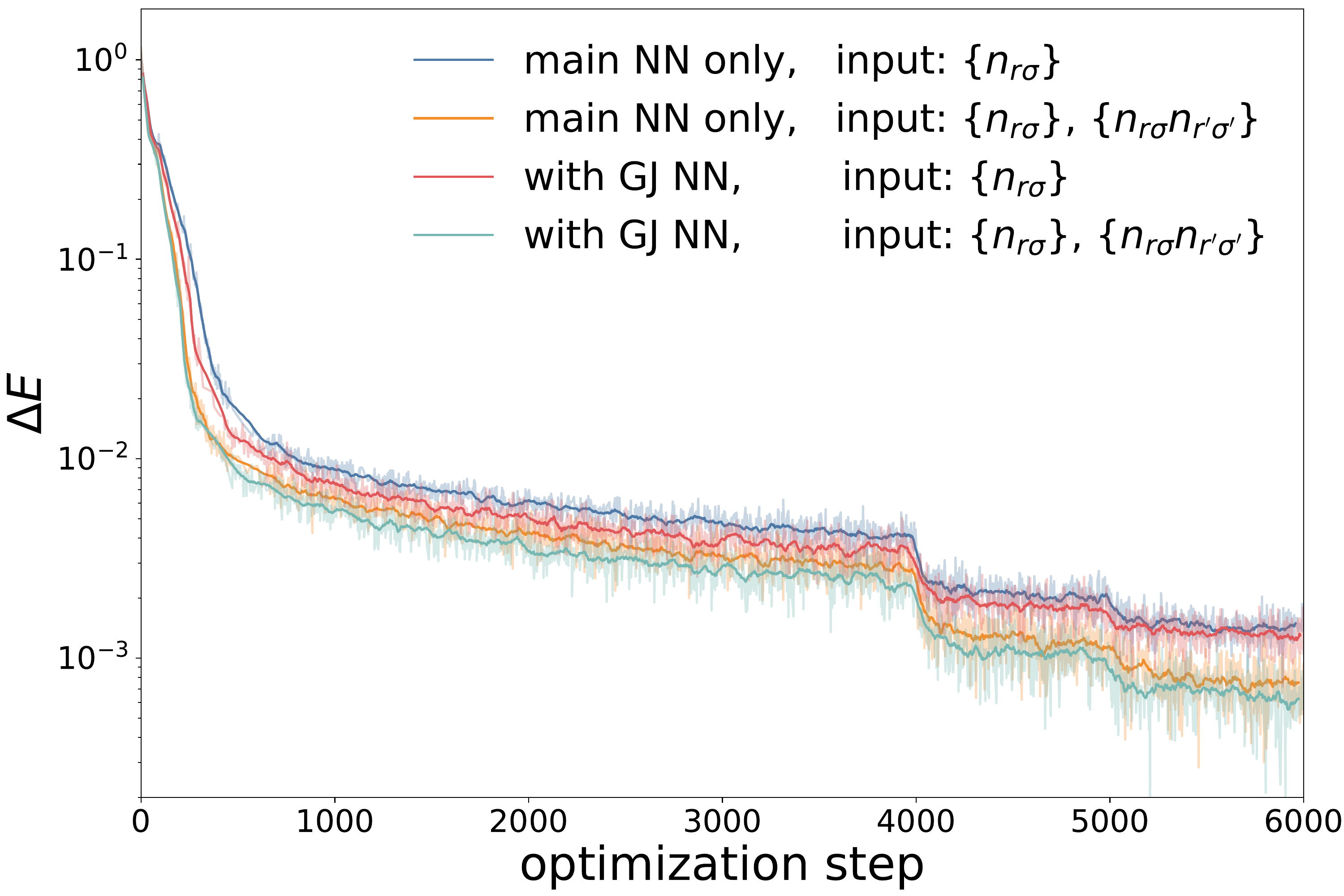}
        \caption{
        Comparison of the optimization processes between different inputs for the main NN, 
        and with and without the use of the GJ-type NN. 
        The parameters and notations are common to those in Fig.~\ref{Fig:mcmc} with $\mu=1$.
}
        \label{Fig:ninj_GJ}
\end{figure}

Next, we investigate how the two NNs introduced in Sec.~\ref{subsec:nn} work in the actual optimization. 
Figure~\ref{Fig:ninj_GJ} shows the optimization processes with and without the input of \{$n_{\bm{r} \sigma} n_{\bm{r}^\prime \sigma^\prime}$\} in the main NN, 
in addition to \{$n_{\bm{r} \sigma}$\}. 
In both cases with and without the use of the GJ-type NN, we find that the additional input accelerates the optimization.
This is surprising since \{$n_{\bm{r} \sigma} n_{\bm{r}^\prime \sigma^\prime}$\} have no extra information 
about the states beyond \{$n_{\bm{r} \sigma}$\}.
The results in Fig.~\ref{Fig:ninj_GJ} also indicate that the use of the GJ-type NN in addition to the main NN accelerates the optimization.

\subsection{Effect of the energy variance} 
\label{subsec:variance}
\begin{figure}[htbp]
        \centering
        \includegraphics[width=\columnwidth]{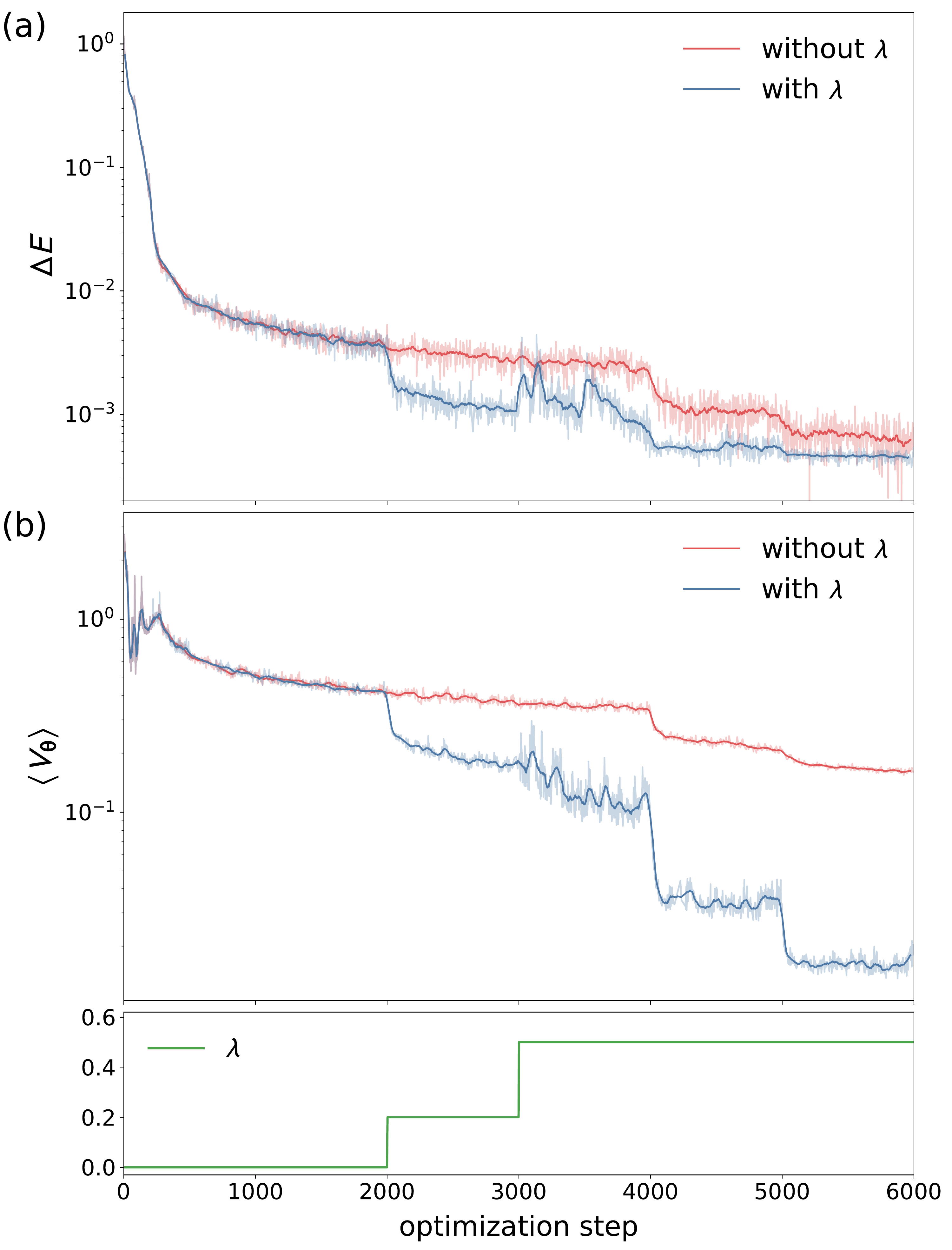}
        \caption{
        Effect of the inclusion of the energy variance $\langle V_{\bm \theta}\rangle$ in the loss function on the optimization process: 
        the step dependences of (a) $\Delta E$ and (b) $\langle V_{\bm \theta}\rangle$. 
        The lower panel shows the schedule of $\lambda$ during the optimization process.
        The other parameters and notations are common to those in Fig.~\ref{Fig:ninj_GJ} 
        with the GJ NN and the input \{$n_{\bm{r}\sigma}$\} and  \{$n_{\bm{r} \sigma} n_{\bm{r}^\prime \sigma^\prime}$\}. 
}
        \label{Fig:lambda}
\end{figure}

\begin{figure}[htbp]
        \centering
        \includegraphics[width=\columnwidth]{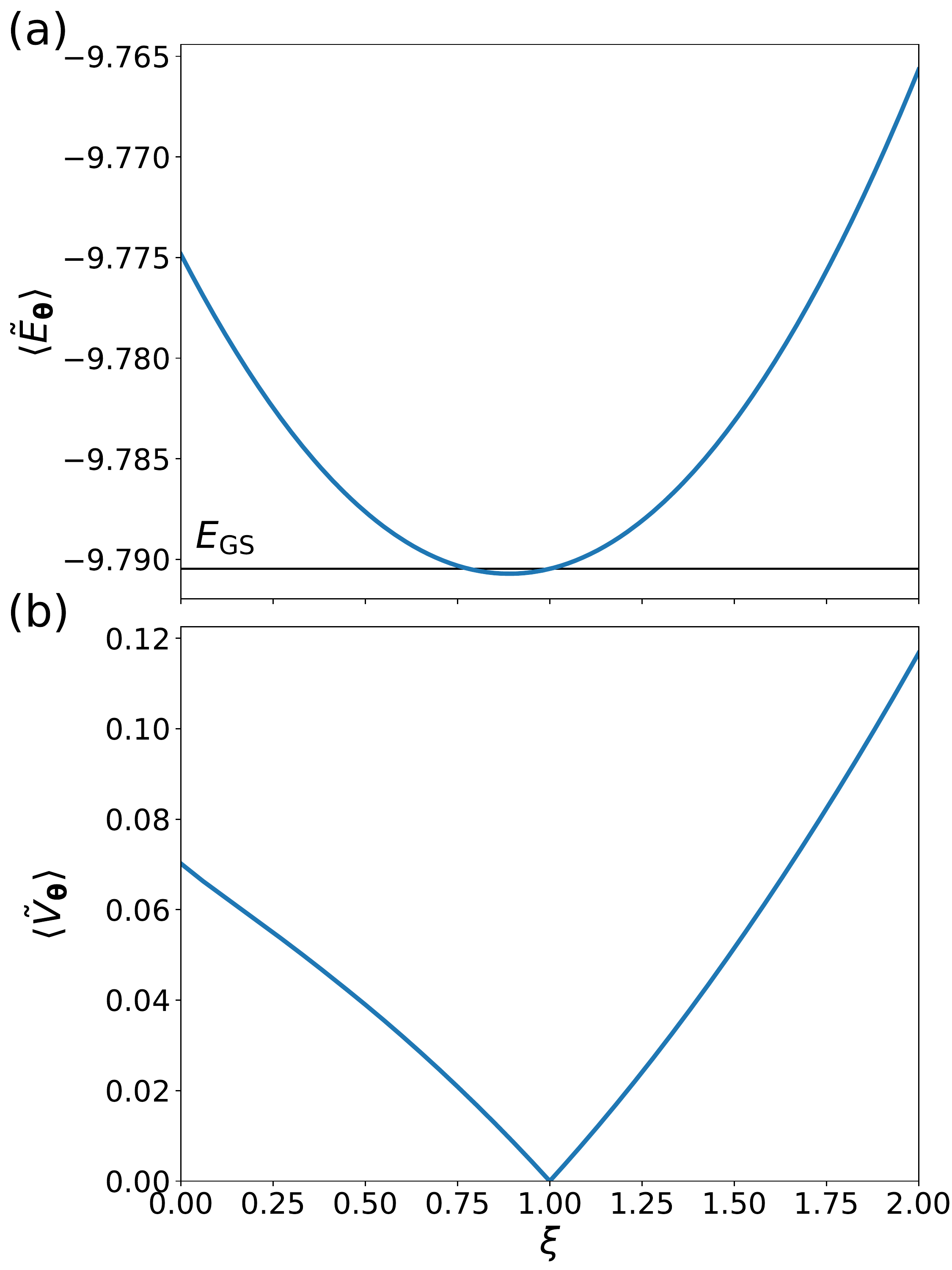}
        \caption{
        $\xi$ dependence of (a) $\langle \tilde{E}_{\bm \theta}\rangle$ and (b) $\langle \tilde{V}_{\bm \theta}\rangle$ calculated by Eq.~\eqref{eq:e_tilde}. 
        We take $\psi_{\bm \theta}(x)$ and $|x\rangle $ at the $1000$th step of the data in Fig.~\ref{Fig:lambda}.
        The horizontal line in (a) represents the true ground state energy.
}
        \label{Fig:landscape}
\end{figure}

Then, we investigate the effect of the inclusion of the energy variance, $\langle V_{\bm \theta}\rangle$ 
in Eq.~\eqref{eq:V}, in the loss function in Eq.~\eqref{eq:L}. 
The results are shown in Fig.~\ref{Fig:lambda}, in comparison with those obtained by the optimization 
without $\langle V_{\bm \theta}\rangle$ 
throughout the optimization. 
Here we vary $\lambda$ in a stepwise manner, as shown in the lower panel of Fig.~\ref{Fig:lambda}(b). 
We find that the introduction of $\lambda=0.2$ at the 2000th step rapidly lowers the energy, 
and at the same time, the energy variance is largely reduced. 
The increase of $\lambda$ to 0.5 at the 3000th step disturbs the optimization temporarily, but
the energy is further lowered in the later steps.
We note that a further increase of $\lambda$ makes the optimization unstable.
In addition, a nonzero $\lambda$ 
from the beginning of the optimization may trap the system into a metastable state 
where only $\langle V_{\bm{\theta}}\rangle$ has a small value.
The results indicate that the appropriate inclusion of the energy variance in the loss function accelerates the optimization process.

In order to understand the reason of this improvement,  
we study the behavior of the energy and the variance around the true ground state. 
We define the linear combination of the wave function using the NN, $\psi_{\bm\theta}(x)$, 
and that for the ground state $\psi_{\rm{GS}}(x) = \langle x | \psi_{\rm GS}\rangle$ as
\begin{align}
\tilde{\psi_{\bm \theta}}(x) = (1-\xi)\psi_{\bm \theta}(x) + \xi\psi_{\rm{GS}}(x). 
\label{eq:psi_r}
\end{align}
Using $|x\rangle$ and $\bm \theta$ at the $1000$th step in the data in Fig.~\ref{Fig:lambda}, 
we compute the average of the energy and the variance for Eq.~\eqref{eq:psi_r} while varying $\xi$ by
\begin{align}
\langle \tilde{O}_{\bm{\theta}} \rangle  = 
\frac{  \sum_{\{x\} \sim |\psi_{\bm{\theta}}(x)|} 
\frac{|\tilde{\psi}_{\bm{\theta}}(x)|}{|\psi_{\bm{\theta}}(x)|} 
|\tilde{\psi}_{\bm{\theta}}(x)|\tilde{O}^{\rm{loc}}_{\bm{\theta}} (x)}
{\sum_{\{x\} \sim|\psi_{\bm{\theta}}(x)|} 
\frac{|\tilde{\psi}_{\bm{\theta}}(x)|}{|\psi_{\bm{\theta}}(x)|} 
|\tilde{\psi}_{\bm{\theta}}(x)|}, 
\label{eq:e_tilde}
\end{align}
where $\tilde{O}$ denotes $\tilde{E}$ or $\tilde{V}$,
and $\tilde{O}^{\rm{loc}}_{\bm{\theta}}(x)$ is calculated by replacing $\psi_{\bm \theta}(x)$ 
with $\tilde{\psi_{\bm \theta}}(x)$ in Eqs.~\eqref{eq:e_loc} and \eqref{eq:V_loc}.

Figure~\ref{Fig:landscape}(a) shows $\langle \tilde{E}_{\bm{\theta}}\rangle$ as a function of $\xi$. 
Although $\langle \tilde{E}_{\bm{\theta}} \rangle = E_{\rm GS}$ at $\xi=1$, 
we find that $\langle \tilde{E}_{\bm{\theta}} \rangle$ takes the minimum slightly below $E_{\rm GS}$ at $\xi\simeq 0.89$. 
While the result appears to violate the variational principle, this is presumably due to the finite number of samples. 
This suggests that the update of $\bm{\theta}$ may fail 
when using only the energy in the loss function.

On the other hand, $\langle\tilde{V}_{\bm{\theta}}\rangle$ takes the minimum value of $\langle\tilde{V}_{\bm{\theta}}\rangle=0$ at exactly $\xi=1$ 
even for a finite number of samples, as shown in Fig.~\ref{Fig:landscape}(b). 
This is because $V_{\bm{\theta}}^{\rm loc}$ in Eq.~\eqref{eq:V_loc} is positive definite and all the local energies $E_{\rm loc}$ 
are equal to $E_{\rm GS}$ in the true ground state, namely, $V_{\bm{\theta}}^{\rm loc}=0$ at $\xi=1$.
This indicates that the optimization can be stabilized by mixing $\langle V_{\bm{\theta}}\rangle$
in the loss function when the system comes close to the ground state, as demonstrated in Fig.~\ref{Fig:lambda}.

\subsection{Irreducible representation}
\label{subsec:irre}
\begin{table}[hbtp]
  \centering
  \begin{tabular}{cccc}
    \hline
    irrep  & $\Delta E$  & $\Delta E_K$ & $\Delta E_U$ \\
    \hline 
    $A_1$  & 0.00048(5)  & 0.000(1)   & 0.00(2)   \\
    $A_2$  & 0.164051(6) & 0.173(2)   & -0.24(1)  \\
    $B_1$  & 0.15091(7)  & 0.167(2)   & -0.27(1)   \\
    $B_2$  & 0.14463(5)  & 0.165(1)   & -0.306(9)   \\
    \hline
  \end{tabular}
  \caption{Errors in the total energy $\Delta E$, the kinetic energy $\Delta E_K$, 
  and the onsite Coulomb energy $\Delta E_U$ for different irreducible representations (irrep). 
  The parameters and calculation conditions are the same as those in Fig.~\ref{Fig:lambda} with the introduction of nonzero $\lambda$. 
  }
  \label{table:irrep}
\end{table}

As discussed in the end of Sec.~\ref{subsec:symm}, we need to choose a proper irreducible representation for the finite-size calculations. 
We list the error of the energy, $\Delta E$, for different irreducible representations in Table~\ref{table:irrep} 
obtained after the optimization in the same condition in Fig.~\ref{Fig:lambda} with the introduction of nonzero $\lambda$. 
The irreducible representation $E$, which has twofold degeneracy, is excluded from the calculations.
We find that the energy for $A_1$ is significantly lower and close to $E_{\rm GS}$,
compared to the other three.
In addition, in Table~\ref{table:irrep}, 
we list the errors in the kinetic energy $\Delta E_K$ and the onsite Coulomb energy $\Delta E_U$
corresponding to the first and second terms in Eq.~\eqref{eq:ham}, respectively.
Except for $A_1$, the kinetic energy remains high, while the onsite Coulomb energy becomes lower than the value in the true ground state. 
This suggests that the inappropriate choice of the irreducible representation fails to optimize 
the kinetic energy presumably due to the wrong sign structure of the wave function.

\subsection{Parameters and samples dependence}
\label{subsec:numdep}

\begin{figure}[htbp]
        \centering
        \includegraphics[width=\columnwidth]{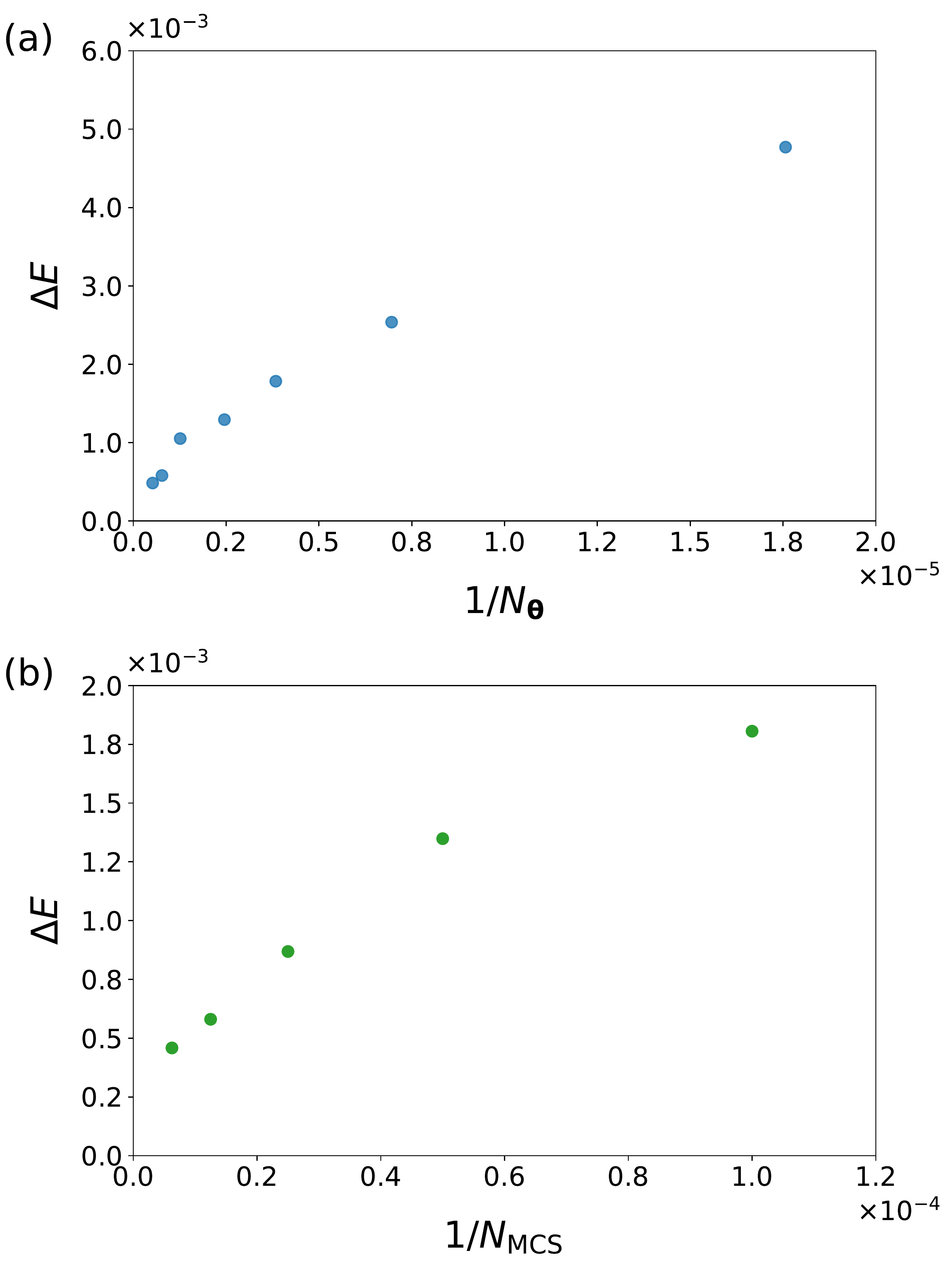}
        \caption{
        (a) $N_{\bm \theta}$ dependence of $\Delta E$.
        We take $N_{\rm MCS}=16000$.
        (b) $N_{\rm MCS}$ dependence of $\Delta E$.
        We take $N_{\bm \theta}=2070120$.
        The computational conditions are the same as those with nonzero $\lambda$ in Fig.~\ref{Fig:lambda}. 
        }
        \label{Fig:samples_dep}
\end{figure}

Figure~\ref{Fig:samples_dep}(a) shows $\Delta E$ while changing the number of NN parameters given by
\begin{align}
N_{\bm \theta} = \sum_{l=0}^{L_{\rm main}-1} D^{\rm main}_lD^{\rm main}_{l+1} + \sum_{ l=0}^{ L_{\rm GJ}-1} D^{\rm GJ}_l D^{\rm GJ}_{l+1},
\end{align}
where $L_{\rm main}$ ($ L_{\rm GJ}$) is the number of layers and $D^{\rm main}_l$ ($D^{\rm GJ}_l$) is the number of neurons 
on the layer $l$ in main (GJ-type) NN.
Here we fix $N_{\rm MCS}$ at $16000$. 
The calculation conditions are the same as those with nonzero $\lambda$ in Sec.~\ref{subsec:variance}.
$\Delta E$ decreases while increasing $N_{\bm \theta}$ and appears to approach 
zero for $N_{\bm \theta}\to\infty$.
This looks consistent with the universal approximation theorem: 
our framework by using the NNs can approximate the true ground state as precisely as possible by increasing the number of parameters, 
when the optimization is performed for sufficiently large samples.

We also plot $\Delta E$ while changing the number of MC samples $N_{\rm MCS}$ in Fig.~\ref{Fig:samples_dep}(b). 
We fix $N_{\bm \theta}$ at $2070120$: 
$L_{\rm main} = 5$, $D_l^{\rm main} = 500$ for $0\leq l \leq 4$ and $D_5^{\rm main} = 40$, 
and $L_{\rm GJ} = 4$, $D_l^{\rm GJ} = 500$ for $0\leq l \leq 3$ and $D_4^{\rm main} = 40$.  
We find that $\Delta E$ monotonically decreases while increasing $N_{\rm MCS}$.
This result indicates that not only the number of NN parameters but also the number of MC samples is relevant to the approximation. 
Contrary to $N_{\bm \theta}$, however, $\Delta E$ appears to converge to a nonzero value for $N_{\rm MCS}\to\infty$, 
suggesting that the capability of the approximation is limited more strongly by $N_{\bm \theta}$ rather than $N_{\rm MCS}$.

\subsection{Benchmark for $N_{\rm site} = 6\times 6$}
\label{subsec:6x6}

\begin{figure}[htbp]
        \centering
        \includegraphics[width=\columnwidth]{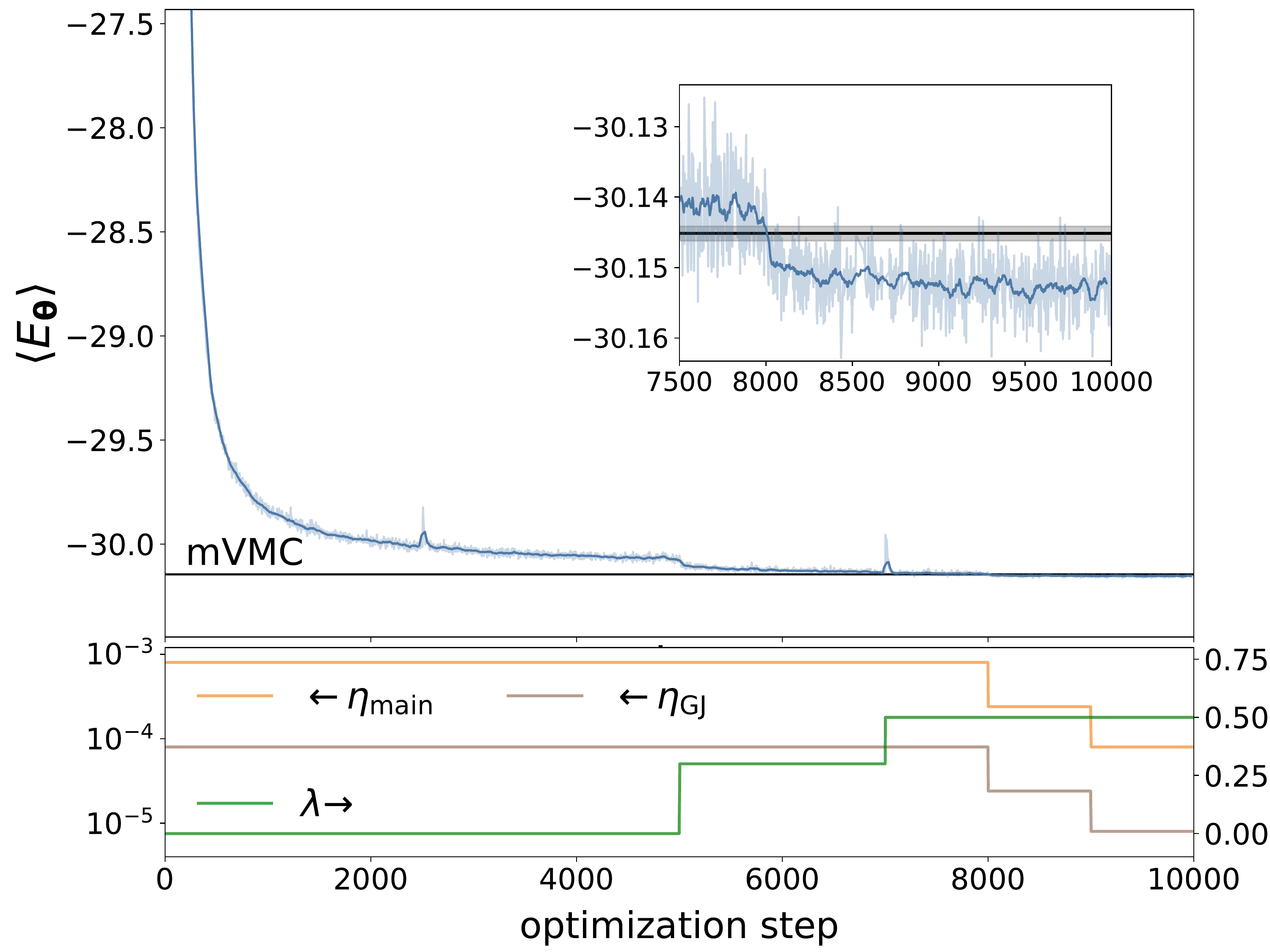}
        \caption{
        Optimization process of energy at $N_{\rm site} = 6\times 6$ and $N = 10$ compared with mVMC. 
        The inset shows an enlarged plot for clarity. 
        The gray shade represents the statistical error of the mVMC data. 
        The lower panel shows the schedules of $\eta_{\rm main}$, $\eta_{\rm GJ}$, and $\lambda$. 
}
        \label{Fig:6x6}
\end{figure}

Finally, we apply our method to the case of $N_{\rm site} = 6 \times 6$ and $N = 10$. 
The result is shown in Fig.~\ref{Fig:6x6}. 
Since the ED is not feasible in this case, we compare the result with that obtained by the mVMC calculations~\cite{MISAWA2019447}. 
We find that the energy by our method is decreased to near the value obtained by the mVMC;
by closely looking, our method achieves a slightly lower energy than the mVMC, as shown in the inset. 
As it was shown that the mVMC can provide comparable or better results compared to the existing methods 
based on NNs~\cite{10.1103/physrevb.96.205152,10.1103/physrevx.10.041026,astrakhantsev2021broken},
our result would prove that the capability of our method is comparable to or better than the other NN-based ones.

\section{concluding remarks}
\label{sec:concludingremarks}

In this paper, we have developed a framework to approximate fermionic many-body wave functions by NNs without using 
the Slater determinant. 
In our framework, the sign change of the wave function associated with the fermionic anticommutation relation is explicitly calculated by defining the order of particles in real space, 
and the rest part of the wave function is optimized by the FCNNs. 
We applied our method to the Hubbard model on the square lattice, and showed that it achieves a lower energy than the mVMC calculation. 
The numerical cost of our framework is $O(N)$-$O(N^2)$, depending on the details of the computational tricks and the input for NNs to be employed. 
This has an advantage over the other NN methods based on the Slater determinant (or Pfaffian) whose numerical cost is $O(N^3)$. 
Thus far, our framework is successful for a modest number of particles, and the application in a larger scale is left for future study; 
it would be crucial to develop further efficient optimization techniques for a larger number of parameters.

In our framework, we have implemented several numerical tricks to stabilize and accelerate the optimization process. 
First, we employed the reweighting method with the use of the MC weight by taking a smaller power than two for the absolute of the wave function, to avoid numerical instabilities by accidentally small weights appearing in the optimization process. 
Second, we limited ourselves to calculate only the representative states,  
which reduces the effective Hilbert space to be calculated and ensures the symmetry of the system for the wave function. 
Lastly, we included the energy variance in the loss function, in addition to the energy. 
We believe that all these tricks would be efficient for a wide class of fermionic systems and applicable to the optimization methods in the conventional VMC studies.

Furthermore, we implemented two FCNNs, the main NN and the GJ-type NN, in the NN architecture.
The main NN represents both the sign and amplitude of the wave function. 
We incorporated not only the number of the particles but also the products of them. 
Meanwhile, the GJ-type NN is a generalization of the GJ factor to incorporate correlation effects. 
We showed that both of them contribute to the improvement of the optimization.
For further improvement, imposing the symmetry of the system on NNs would give better results~\cite{Nomura_2021}.
In addition, it is interesting to incorporate various NN architectures developed in the machine learning community.

\begin{acknowledgments}
The authors thank T. A. Bojesen, T. Misawa, Y. Nomura, and N. Yoshioka for fruitful discussions.
They acknowledge the use of open-source software $\mathcal{H}\Phi$~\cite{KAWAMURA2017180} for the ED calculations. 
This research was conducted using the SGI Rackable C2112-4GP3/C1102-GP8 (Reedbush-U/H/L) in the Information Technology Center, the University of Tokyo. 
The ED and mVMC calculations were performed on the facilities of the Supercomputer Center, the Institute for Solid State Physics, the University of Tokyo.
This work was supported by JST CREST (JP-MJCR18T2) and JSPS KAKENHI Grant No.~20H00122.
\end{acknowledgments}

\appendix
\section{Derivative of loss function}
\label{app:lf}

Equation~\eqref{eq:o} is derived as \begin{widetext}
\begin{align}
\nabla_{\bm{\theta}} \langle O_{\bm{\theta}} \rangle &= 
\nabla_{\bm{\theta}} \frac{\sum_x |\psi_{\bm{\theta}}(x)|^2 O^{\rm{loc}}_{\bm{\theta}}(x)}{\sum_x |\psi_{\bm{\theta}}(x)|^2} \notag \\  
&=\nabla_{\bm{\theta}} \frac{\sum_x |\psi_{\bm{\theta}}(x)|^\mu |\psi_{\bm{\theta}}(x)|^{2-\mu} O^{\rm{loc}}_{\bm{\theta}}(x)}{\sum_x |\psi_{\bm{\theta}}(x)|^2} \notag \\  
&=
\frac{\sum_x \left[ \mu |\psi_{\bm{\theta}}(x)| \left(\nabla_{\bm{\theta}} |\psi_{\bm{\theta}}(x)|\right)  O^{\rm{loc}}_{\bm{\theta}}(x) + 
    |\psi_{\bm{\theta}}(x)|^\mu \nabla_{\bm{\theta}} \left(|\psi_{\bm{\theta}}(x)|^{2-\mu}O^{\rm{loc}}_{\bm{\theta}}(x)  \right) \right]}
    {\sum_x |\psi_{\bm{\theta}}(x)|^2} \notag \\
    &\quad- 
\frac{\sum_x |\psi_{\bm{\theta}}(x)|^2 O^{\rm{loc}}_{\bm{\theta}}(x)}
     {\sum_x |\psi_{\bm{\theta}}(x)|^2}
\frac{\sum_x 2|\psi_{\bm{\theta}}(x)| \left( \nabla_{\bm{\theta}} |\psi_{\bm{\theta}}(x)| \right)}
     {\sum_x |\psi_{\bm{\theta}}(x)|^2} \notag \\  
&
\stackrel{\mathrm{(*)}}{=}
\frac{\sum_x \left [|\psi_{\bm{\theta}}(x)|^2 \left(\mu O^{\rm{loc}}_{\bm{\theta}}(x) - 2\langle O_{\bm{\theta}} \rangle \right) \nabla_{\bm{\theta}} \log (|\psi_{\bm{\theta}}(x)|) +
     |\psi_{\bm{\theta}}(x)|^\mu \nabla_{\bm{\theta}} \left(|\psi_{\bm{\theta}}(x)|^{2-\mu}O^{\rm{loc}}_{\bm{\theta}}(x) \right) \right]
     }
     {\sum_x |\psi_{\bm{\theta}}(x)|^2} \notag \\
&=
\frac{\sum_x |\psi_{\bm{\theta}}(x)|^\mu \left[ \left( O^{\rm{loc}}_{\bm{\theta}}(x) - \frac{2}{\mu}\langle O_{\bm{\theta}} \rangle \right) 
|\psi_{\bm{\theta}}(x)|^{2-\mu} \nabla_{\bm{\theta}} \log (|\psi_{\bm{\theta}}(x)|^\mu) +
      \nabla_{\bm{\theta}} \left(|\psi_{\bm{\theta}}(x)|^{2-\mu}O^{\rm{loc}}_{\bm{\theta}}(x)\right) \right]
     }
     {\sum_x |\psi_{\bm{\theta}}(x)|^\mu |\psi_{\bm{\theta}}(x)|^{2-\mu}} \notag \\
& \approx
\frac{ 
     \sum_{\{x\} \sim |\psi_{\bm{\theta}}(x)|^\mu} \left[
        \left( O^{\rm{loc}}_{\bm{\theta}}(x)-\frac{2}{\mu}\langle O_{\bm{\theta}} \rangle \right) 
        |\psi_{\bm{\theta}}(x)|^{(2-\mu)} \nabla_{\bm{\theta}} \log (|\psi_{\bm{\theta}}(x)|^\mu) 
        + \nabla_{\bm{\theta}} \left(|\psi_{\bm{\theta}}(x)|^{(2-\mu)}O^{\rm{loc}}_{\bm{\theta}} (x) \right)
    \right]
}{
    \sum_{\{x\} \sim|\psi_{\bm{\theta}}(x)|^\mu} |\psi_{\bm{\theta}}(x)|^{(2-\mu)} 
}.
\end{align}
\end{widetext}
We use the relation 
$\nabla_{\bm{\theta}}|\psi_{\bm{\theta}}(x)| = |\psi_{\bm{\theta}}(x)|\nabla_{\bm{\theta}} \log|\psi_{\bm{\theta}}(x)|$ 
from the fourth to fifth line denoted by ($*$).

\bibliography{main}

\end{document}